\documentstyle[prl,twocolumn,aps,epsfig,epsf]{revtex}
\begin{document}
\draft
\title{Quantum Zeno effect and parametric resonance in mesoscopic physics}
\author{G. Hackenbroich$^1$, B. Rosenow$^2$, and H. A. Weidenm\"uller$^2$}
\address{$^1$Universit\"at GH Essen, Fachbereich 7, 45117 Essen, Germany}
\address{$^2$Max-Planck-Institut f\"ur Kernphysik, D-69029 Heidelberg}
\date{\today}
\maketitle
\begin{abstract}
  As a realization of the quantum Zeno effect, we consider electron
  tunneling between two quantum dots with one of the dots coupled to a
  quantum point contact detector. The coupling leads to decoherence
  and to the suppression of tunneling. When the detector is driven
  with an ac voltage, a parametric resonance occurs which strongly
  counteracts decoherence. We propose a novel experiment with which
  it is possible to observe both the quantum Zeno effect and the
  parametric resonance in electric transport.
\end{abstract}
\narrowtext \tighten 

\vspace*{-0.3cm}
The interaction of a quantum system with a macroscopic measurement
device is known \cite{Wheeler,Zurek} to generate decoherence. The
frequent repetition of a decohering measurement leads to a striking
phenomenon known as the quantum Zeno effect \cite{Misra}: The
suppression of transitions between quantum states. The standard
example is a two--level system with a tunneling transition between the
two levels. For small times, the probability to tunnel out of one of
the two levels is $\sim t^2$. With a device that projects the system
onto that same level, $N$ repeated measurements yield the reduced
probability $\sim N (t/N)^2$. The suppression of tunneling in bound
systems has its parallel in systems with unstable states. Here, the
quantum Zeno effect predicts the suppression of decay and the
enhancement of the lifetime of unstable states.

Despite considerable theoretical work on the quantum Zeno effect there is only
little experimental proof for it. An experimental test using an induced
hyperfine transition of Be ions \cite{Itano} has been reported.  Experiments
on optical transitions \cite{Gagen} or atomic Bragg scattering \cite{Dyrting}
have been proposed. However, the observation of the quantum Zeno effect in
some of these experiments is hampered by other sources of decoherence not
related to the measurement.  Further experimental evidence for the existence
of the quantum Zeno effect is clearly desirable.  Gurvitz \cite{Gurvitz2}
first pointed out that the quantum Zeno effect may be observed in
semiconductor microstructures.  He studied theoretically the tunneling of an
electric charge between two weakly coupled quantum dots. A quantum point
contact (QPC) located in the vicinity of one of the dots served as a
non--invasive detector for the charge on that dot.  As expected from the
quantum Zeno effect, Gurvitz found that the coupling to the QPC suppresses the
tunneling oscillations between the two dots.

In this Letter we use a microscopic theory to study the quantum Zeno effect in
coupled quantum dots. We propose an arrangement which is within reach of
present-day experimental techniques, and through which the quantum Zeno effect
can be investigated. Our work is motivated by the first experimental
demonstration of complementarity and controlled dephasing in a which-path
semiconductor device by Buks {\em et al.} \cite{Buks}. The observation of
interference in an Aharonov--Bohm interferometer with a quantum dot (QD)
embedded in one of its arms showed that electrons pass coherently through the
device and the QD. A QPC measuring the electric charge in the QD led to
dephasing. Several theoretical explanations of the experimental results have
been given \cite{Buks,Gurvitz1,Levi97,AlWiMe97}. Our  work goes considerably
beyond that of Ref.~\cite{Gurvitz2}. Indeed,  we identify several
novel 
aspects of
the quantum Zeno effect in coupled quantum dots. First, we show that the
application of an AC voltage with frequency $\omega$ across the QPC leads to
parametric resonance and to a strong reduction of decoherence. The resonance
occurs when $\omega$ equals twice the frequency $\omega_0$ of the internal
charge oscillations in the double--dot system. Second, the power
spectrum of 
the QPC
displays a peak which is a clear signal for the quantum Zeno effect.  Third,
we propose a new transport experiment with the two quantum dots in a parallel
circuit and each dot coupled to external leads. With current flowing into the
first dot, we calculate the branching ratio of the current transmitted through
the second dot and that through the first dot.  Measurements with the
QPC-detector necessarily induce an energy exchange between the dots and the
QPC. Therefore, the coupling to the QPC induces a correction to the branching
ratio which has both an elastic and an inelastic contribution. This correction
is proportional to the decoherence rate. A measurement of the branching ratio
thus provides a direct signature of dephasing, and of the quantum Zeno effect
in the two coupled quantum dots.

We deal first with a simple system: Two coupled quantum dots without
coupling to external leads and occupied by a single (excess) electron. 
We consider only one energy level in each dot and assume that both
levels are degenerate. Let $E_0$ denote the energy of each level and
$\Omega_0/2$ with $\Omega_0 = \hbar \omega_0$ the coupling matrix
element between the two dots. The lower dot interacts with a
QPC, which continuously measures the electron position. 
The QPC is 
modeled as a single--channel device which is
symmetric with respect to the lower dot. The relation of this simple
model to a realistic dot system is briefly discussed later. In order
to derive a master equation for the density matrix $\rho_{\rm   dot}$
of the two quantum dots, we calculate the total density matrix of dots
plus QPC and then trace out the QPC--variables. The total density
matrix is obtained from a scattering approach which generalizes a
method first introduced in Ref.~\cite{Buks}. The scattering matrix
$S_{\rm QPC}$ through the QPC has dimension two and depends on the
location of the electron. If the electron is on the lower [upper] dot,
we write $S$ in the form $S_{\rm QPC}=\exp(i \theta_l \tau_x)$
[$S_{\rm QPC}=\exp(i \theta_u \tau_x)$], respectively. Here, $\tau_x$
is a Pauli spin matrix. In both cases we have suppressed a global
phase, and we have used time--reversal symmetry and the above--mentioned
symmetry of the QPC. Both cases can be combined in writing the
two--particle scattering matrix as $S_{\sigma \sigma^\prime} =
\delta_{\sigma \sigma^\prime} [\delta_{\sigma l} e^{i \theta_l \tau_x}
+ \delta_{\sigma u} e^{i \theta_u \tau_x} ]$, where $\sigma$ and
$\sigma^\prime$ both stand for either quantum dot labeled $l$ and
$u$, respectively. With $\rho^{(0)}= \rho_{\rm dot}^{(0)} \, \,
\rho_{\rm QPC}^{(0)}$ the density matrix of the total system prior to
the passage of an electron through the QPC, the density matrix after
scattering through the QPC is given by $\rho = S \rho^{(0)} S^\dagger$. 
The reduced density matrix of the two dots is obtained by tracing over
the QPC-variables and given by
\begin{equation}
\rho_{\rm dot} = {\rm Tr}_{\rm QPC} \rho = \rho_{\rm dot}^{(0)}\! - {1
    \over 2} (\Delta \theta)^2 \! \left( \begin{array}{cc} 0\! & \rho_{{\rm
      dot},12}^{(0)} \\ \rho_{{\rm dot},21}^{(0)} & \!0 \end{array}
\right).
\label{event}
\end{equation}
We have used $\Delta \theta = \theta_u \!\!- \theta_l$. Assuming that
the coupling between the lower dot and the QPC is weak, we have
expanded $\cos (\Delta \theta) \approx 1-(\Delta \theta)^2/2$. The
term $(\Delta \theta)^2$ can be expressed in terms of the transmission
coefficients ${\cal T}_l,{\cal T}_u$ \cite{Buks} so that $(\Delta
\theta)^2 = (\Delta {\cal T})^2 / [4 {\cal T} (1-{\cal T})]$ where
$\Delta {\cal T} = {\cal T}_u - {\cal T}_l$ and ${\cal T} = ({\cal
  T}_u + {\cal T}_l)/2$. We simplify the notation by writing the last
term in Eq.~(\ref{event}) as $ - (\Delta \theta)^2 \Sigma/2$. We note
that with $\mu$ the voltage drop across the QPC, the time between two
scattering events in the QPC is given by $\Delta t = h /(2 e
\mu)$. During this time, the dynamics of the two--dot system is
governed by the tunneling Hamiltonian $\Omega_0 \sigma_x
/2$. Combining this fact with Eq.~(\ref{event}), we obtain the master
equation
\begin{equation}
  {d \rho_{\rm dot} \over d t}\! = - { (\Delta {\cal T})^2 \over 8
    {\cal T} (1-\!\!{\cal T}) } {e \mu \over \pi \hbar} \Sigma -{i
    \Omega_0 \over 2 \hbar} [\sigma_x, \rho_{\rm dot}].
\label{master}
\end{equation}
A master equation for the double--dot system has previously been
derived by Gurvitz \cite{Gurvitz2} using a different method. We note
that the factor $1-{\cal T}$ is missing in Gurvitz' result.

Since $\rho_{\rm dot}^{} = \rho^\dagger_{\rm dot}$ and ${\rm Tr} \rho =1$, we
can parameterize $\rho_{\rm dot}$ by the two quantities $a=1/2-\rho_{{\rm
    dot},11}$, $b=\rho_{{\rm dot},12}$. Substitution into Eq.~(\ref{master})
yields the equations of motion $(d a)/ (d t) = \omega_0 {\rm Im} b$ and
\begin{equation}
{d^2 a \over dt^2} +  { (\Delta {\cal T})^2 \over 8 {\cal T} 
(1-\!\!{\cal T}) } {e \mu \over \pi \hbar}  {d a \over dt} +\omega_0^2 a =0 .
\label{osci}
\end{equation}
For a time--independent voltage drop $\mu=\mu_0$ both $a$ and $b$
display damped oscillations $\sim \exp[-\kappa t] \cos
\sqrt{\omega_0^2\! -\! \kappa^2} t$ with the damping constant $\kappa
= (\Delta \theta)^2 e \mu / ( 4 \pi \hbar)$. The exponential
suppression of oscillations in $a$ clearly demonstrates the quantum
Zeno effect. The off--diagonal elements of $\rho$ vanish for large
times in agreement with the standard picture of decoherence, and the
density matrix reduces to the random statistical ensemble $\rho_{\rm dot}
\rightarrow {\rm diag}(1/2,1/2)$. The charge oscillations in the
double--dot system modulate the current in the QPC and, hence, modify
its power spectrum. They cause a peak with FWHM $2 \kappa$ centered at
the shifted frequency $\omega_0 \sqrt{1 - \kappa^2/\omega_0^2}$. This
peak will be present in addition to standard shot noise. Both location
and width of this peak are clear signatures of the quantum Zeno effect. 
The experimental investigation of these features would amount to a
time--resolved study of the quantum--mechanical measurement process
and would be of considerable interest.

Interesting new physical aspects arise if $\mu$ has an AC-component,
$\mu(t) = \mu_0-\mu_1 \sin \omega t$ where $\mu_0,\mu_1 \ge 0$ and
$\mu_1 \le \mu_0$. According to Eq.~(\ref{osci}) this corresponds to a
harmonic oscillator with an {\em oscillatory} damping constant. Using
a simple ansatz for $a(t)$ and neglecting terms of order ${\cal O}
(\mu^2)$, one is led to an equation of the Mathieu type which is known
\cite{landau} to display parametric resonance close to the frequencies
$\omega = 2 \omega_0 /n$ where n is a positive integer. Parametric
resonance is most pronounced for $\omega \approx 2 \omega_0$. The
damping near the resonance is strongly reduced, $\kappa =  { (\Delta
  {\cal T})^2 \over 8 {\cal T} (1-\!\!{\cal T}) } {e  \over  2 \pi
  \hbar} (\mu_0-{1 \over 2} \mu_1)$. The resulting time evolution of
$a$ near resonance is illustrated in Fig.~1 and compared with the case
where $\mu$ is time--independent. The resonance condition $\omega = 2
\omega_0$ is interpreted as follows: The position of the electron is
not measured when $\mu$ is close to zero. The electron uses this time
to tunnel from one dot to the other. \vspace*{-0.2cm}

\begin{figure}
\epsfig{file=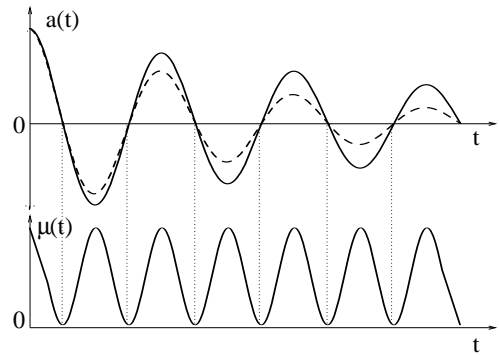,width=4.7cm,angle=270}
\vspace*{.2cm}
\caption{Parametric resonance in the double-dot system coupled to a
  QPC-detector. The upper part shows the oscillations of the diagonal
  elements of the density matrix for constant voltage (dashed curve)
  and for a time-dependent voltage as shown in the lower part (full
  curve). Damping of the oscillations is reduced by a factor two.}
\label{fig1}
\end{figure}

\vspace*{-0.1cm}
We now ask: How does the quantum Zeno effect manifest itself in a
transport experiment? We consider the arrangement shown in Fig.~2
with the two dots in parallel and each dot coupled to two external
leads. For simplicity we consider only one transverse channel in each
lead labeled $c$. The quantum dots are assumed to be in the resonant
tunneling regime close to a Coulomb blockade resonance so that it is
sufficient to consider only a single level in each dot. Both levels
are assumed to have the same energy $E_0$ and width $\Gamma$. The QPC
detector is described in terms of plane waves with energy $\epsilon_k$
and mean density $\rho_F$ which are scattered from a spatially local
potential with Fourier components $U_{k k^\prime}$. To model the
capacitive coupling of the QPC with the lower dot, we write the
Hamiltonian for the QPC plus interaction as
\begin{eqnarray}  H'  = \sum_{k} \epsilon^{}_k b_k^\dagger 
b_{k}^{} + \sum_{k,k^\prime} ( U_{k k^\prime} + V_{k k^\prime} 
d^\dagger_l d^{}_l) b_k^\dagger b_{k^\prime}^{} \ .
\label{Ham}
\end{eqnarray}
Here $b^{\dagger}_k$ and $d^{\dagger}_{\sigma}$, $\sigma={l,u}$ 
denote the creation operators for the QPC states and for the states on
either QD, respectively. We note that the interaction vanishes for an
electron on the upper dot (see Fig.~2).\\
\hspace*{.8cm}
\epsfig{file=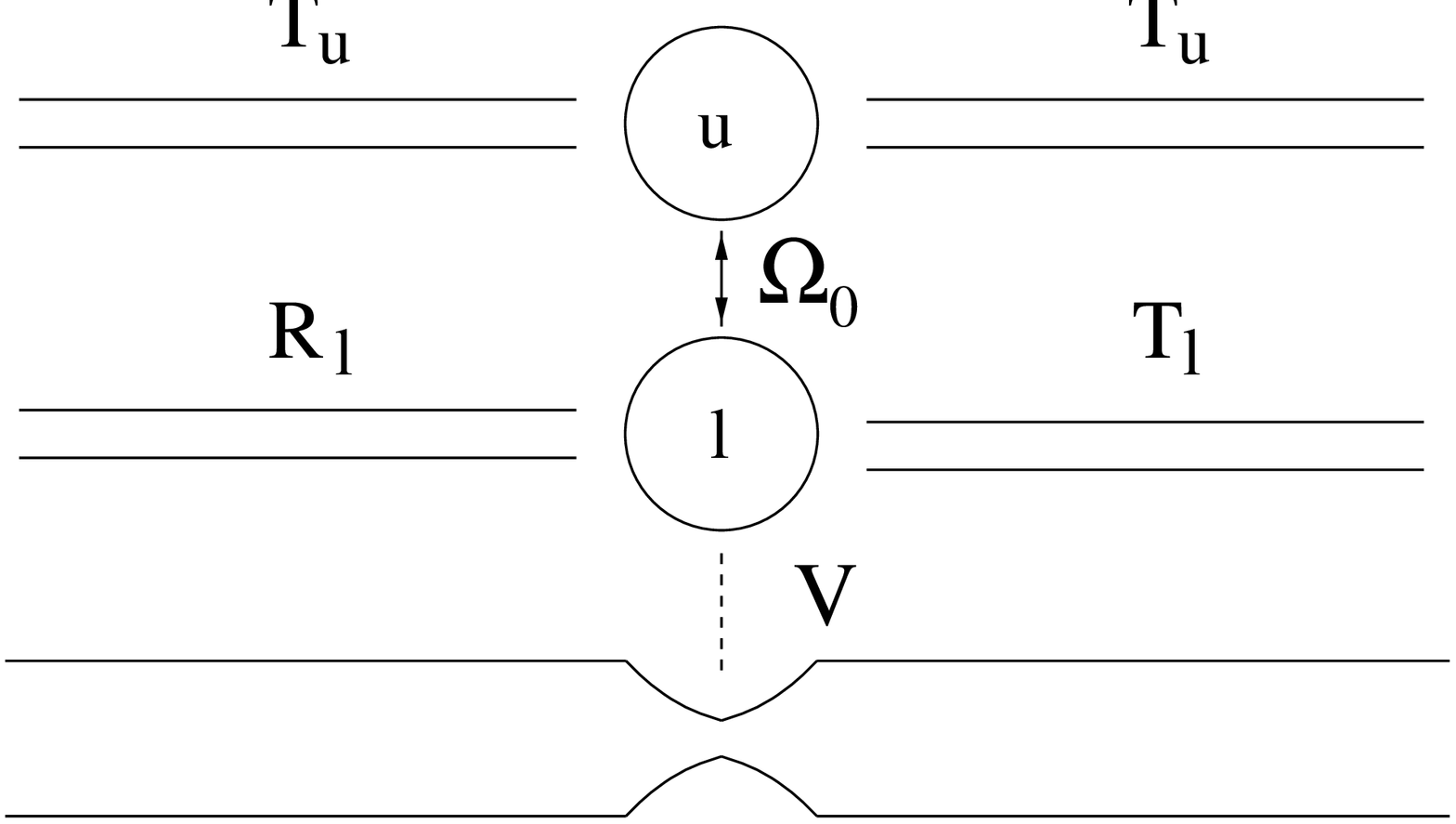,width=3.7cm,angle=270}
\begin{figure}
\epsfig{file=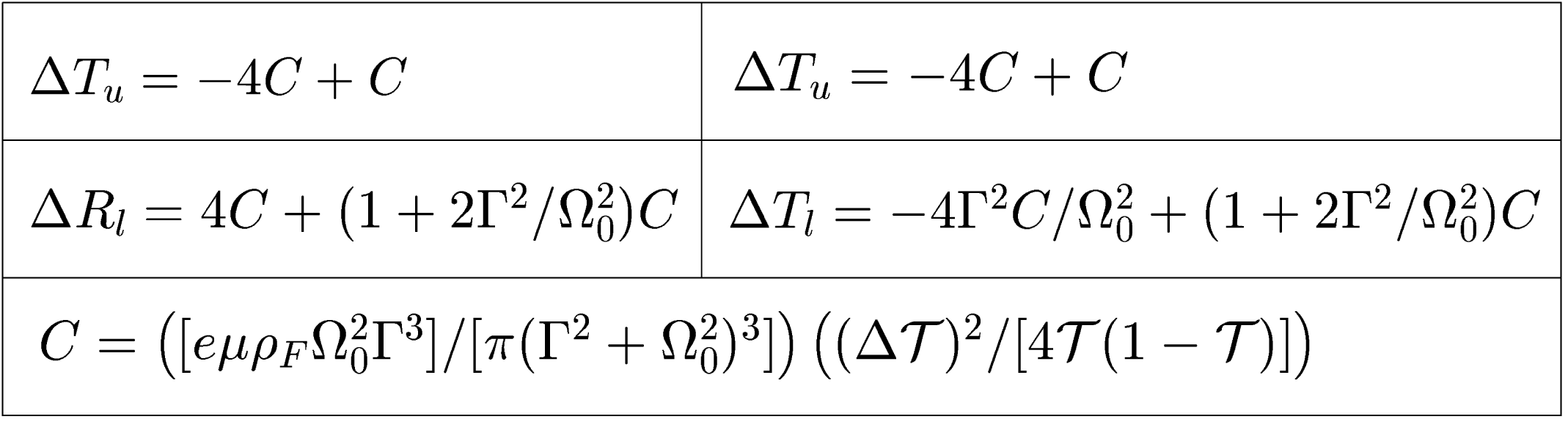,width=2.1cm,angle=270}
\vspace*{.3cm}
\caption{Upper part: double--dot system coupled to a quantum point contact. 
Lower part: the table shows the change of transmission and reflection
  coefficients of the double--dot system due to coupling with the QPC
  detector. For each coefficient the first term gives the elastic and
  the second one the inelastic contribution.}
\label{fig2}
\end{figure}

To find the transmission through the system, we calculate the {\em
  two--particle scattering amplitude} $S_{\sigma c,\sigma^{\prime}
  c^{\prime}; k k^{\prime}}$ for the joint transition between channels
$\sigma c$ and $\sigma^{\prime} c^{\prime}$ in the leads and the
states $k$ and $k^{\prime}$ in the detector from the
Lippmann--Schwinger equation. This yields
\begin{equation}
  S_{\sigma c,\sigma^\prime c^\prime; k k^\prime} = \delta_{ \sigma
    \sigma^\prime } \delta_{c c^\prime} \delta_{k k^\prime} \! \!- \!
  2 \pi i \gamma_{\sigma c} \gamma_{\sigma^\prime c^\prime} G_{\sigma
    \sigma^\prime, k k^\prime} \! .
\label{scatt2}
\end{equation}
Here, $G$ is the two-particle Green function for the joint transition
through the dots and the QPC, and $\gamma_{\sigma, c}$ denotes the
partial width amplitude for the decay of a resonance from the dot
$\sigma$ into the channel $c$. The explicit form of $G$ is given in
terms of the matrix elements 
\begin{equation} 
\label{Green}
[G^{-1}]_{k k^\prime} =  \left(\! \begin{array}{cc}
  G_0^{-1} \delta_{k k^\prime}\!-\!U_{k k^\prime}&  \Omega_0 /2\;
  \delta_{k k^\prime}\\ 
\Omega_0/ 2\; \delta_{k k^\prime} & G_0^{-1} \delta_{k
k^\prime}\!-\!U_{k k^\prime}\!-\!V_{k k^\prime} \end{array}\! \right)
\end{equation}
of its inverse where the $\sigma$--dependence has been displayed
explicitly. Here $G_0^{-1}= E-E_0 -\epsilon_k + i \Gamma/2$ is the
inverse propagator of the single Breit--Wigner resonance, and $E$ is
the total energy of the two incoming particles. We have assumed that
all partial width amplitudes $\gamma_{\sigma c}\equiv \gamma$ are
identical and have used the relation $\Gamma =4 \pi \gamma \gamma^*$. 
We note that for $V = 0$, $G$ simply reduces to the product of the
unit matrix $\delta_{k k^\prime}$ and the two coupled Breit--Wigner
resonances for the double--dot system. Aside from the appearance of
the two--particle Green function $G$, the expression (\ref{scatt2})
differs from the scattering matrix for a double--dot system without
detector by allowing for energy exchange between the dots and the
detector. This manifests itself in the appearance of a factor
$\delta_{ \epsilon_{c^{}} \! + \! \epsilon_{k^{}}, \epsilon_{c^\prime}
  + \epsilon_{k^\prime}}$ multiplying the S-matrix and indicating
overall 
energy conservation. 
Our expression for $[G^{-1}]_{k k^\prime}$ includes the effects of the
interaction (\ref{Ham}) but not many--body effects due to the presence
of other electrons in the leads. To investigate such effects we also
calculated the two--particle scattering amplitude using the
LSZ--formalism \cite{LSZ}. The most important modification to
Eq.~(\ref{scatt2}) comes from the Pauli principle which reduces the
phase space available for scattering and leads to a suppression of
dephasing, cf. our remarks at the end of this paper and a future
publication \cite{HRW}.

We restrict ourselves to constant scattering potentials $U_{k
  k^\prime} \equiv U$ and $V_{k k^\prime} \equiv V$ and calculate the
Green function from Eq.~(\ref{Green}) by expanding to all orders in $V$
and resumming the resulting series. We find two types of contributions
to $G$. The first contribution is independent of $V$ and describes
independent elastic scattering through the QPC and the dots. This term
is not given here. The second contribution represents the connected
part and involves energy exchange $\Omega$ between the dots and the
QPC. It is a complicated function of the energy $E$ but the essential
features are displayed at $E=E_0 + \epsilon_k$,    
\begin{eqnarray}  G^{\rm conn}_{k k^\prime} (\Omega)\! & =\! &  
{4\over \Gamma^2 + \Omega_0^2}\! \ {-V\over F_U F_{U \!+ \!V}} \
{\delta_{\epsilon_k, \epsilon_{k^\prime} +\Omega} \over (2 \Omega + i
  \Gamma )^2 - \Omega_0^2 }
\nonumber\\
& & \hspace*{-1.7cm} \hspace*{1.7cm}
\times \left( \begin{array}{cc} \Omega_0^2 &-i  \Omega_0 \Gamma \\
\!\!-\Omega_0 ( 2 \Omega + i \Gamma) & i \Gamma ( 2\Omega + i \Gamma)\!\!
\end{array} \right) \ .
\label{inter}
\end{eqnarray}
Here, $F_U=1+2 \pi i U \rho_{F}$. The $\Omega$--dependent
prefactor effectively limits inelastic processes to an interval of
width $\Gamma$ around $E_0 + \epsilon_k$. The energy exchange is
essential to  ensure the unitarity of the S--matrix. Moreover, it
allows for a position measurement of the dot electron without
violating the Heisenberg uncertainty relation.

To calculate the single--particle transmission and reflection
coefficients through the double--dot system we trace over the degrees
of freedom of the QPC. We emphasize that we add the two--particle
scattering probabilities and not the two--particle scattering 
amplitudes since the paths of electrons going through
the QPC can be observed in principle. In the case $V=0$ all
scattering processes are elastic and the transmission and reflection
coefficients at resonance $E=E_0 + \epsilon_k$ are $ T^{(0)}_u =
\Omega_0^2 \Gamma^2 (\Gamma^2 + \Omega_0^2)^{-2}$, $T^{(0)}_l =
\Gamma^4 (\Gamma^2 + \Omega_0^2)^{-2}$, and $R^{(0)}_l = \Omega_0^4
(\Gamma^2 + \Omega_0^2)^{-2}$. The branching ratio is $T^{(0)}_u
/T^{(0)}_l= \Omega_0^2 / \Gamma^2$. We note that the branching ratio
has no oscillatory dependence on $\Omega_0 / \Gamma$ as one might
expect when naively applying the idea of time--dependent oscillations
between the two quantum dots to a transport problem.

Corrections to the transmission coefficients due to the interaction $V$
are calculated in a weak--coupling expansion to second order in
$V$. This is the appropriate limit realized experimentally \cite{Buks}. 
In this limit the application of a drain source voltage $\mu$ across
the QPC is equivalent to the simultaneous scattering of $2 e \mu
\rho_F$ particles in different {\em longitudinal} QPC modes. The total
effect of these particles is obtained by multiplying the result for
one QPC--particle with the number of longitudinal modes.

The corrections to the transmission and reflection coefficients due to
$V$ are collected in Fig.~2 and arise both from coherent (elastic)
and incoherent (inelastic) scattering. When the quantum dots are
embedded in an Aharonov--Bohm interferometer only the coherent part of 
transport contributes to the flux--dependent current oscillations. 
However, in the setup depicted in Fig.~2,  one measures the total
current. The results show that measurements with the QPC detector have
a twofold effect: (i) They suppress tunneling from the feeding lead
into the lower dot and (ii) they suppress tunneling from the lower
into the upper dot. Observation (i) follows from the {\em increase} in
reflection, and (ii) from the {\em decrease} of the branching ratio 
\begin{eqnarray}
  {T_{u}\over T_{l}}={\Omega_0^2\over \Gamma^2} \left[ 1
  - {e\mu\over \pi \Gamma}{(\Delta {\cal T})^2 \over {4 \cal T} (1-
    {\cal T})}\right]\ .
\label{branching}
\end{eqnarray}
Both effects (i), (ii) have an obvious interpretation as manifestations of the
quantum Zeno effect. We note that the coefficients given in Tab.~1 depend on
our choice for the scattering amplitudes $\gamma_{\sigma c}$ at the junctions
in the circuit of Fig.~2. However, one universally finds an increase of
reflection and a decrease of the branching ratio. Moreover, 
the coefficient given in square brackets in Eq.~(\ref{branching}) is
independent of the $\gamma_{\sigma c}$'s.  We add that the second term in the
square bracket is up to a factor $\Gamma / (4 \hbar)$ the damping constant
found for the isolated double--dot system. The appearance of the damping
constant in the branching ratio shows that the parametric resonance discussed
above for isolated dots can also be observed in a transport experiment.

An interesting special case of the S--matrix (\ref{scatt2}) is
obtained for $\Omega_0 = 0$ when the two dots are completely isolated
from each other. In this limit we can study how the interaction with
the QPC reduces the transmission through a single dot. This
corresponds to the reduction of the Aharonov--Bohm contrast in the
experiment of Ref.\cite{Buks}. For the modulus of the elastic
transmission amplitude at resonance, we find
\begin{equation}
|t_{el}| =  
\sqrt{T_{el}}=1- {e \mu \tilde{\rho}_F \over \pi \Gamma} {(\Delta 
{\cal T})^2
\over {2 \cal T} (1- {\cal T})}
\label{Buks}
\end{equation}
which is up to a factor of $1/2$ in agreement with previous
calculations \cite{Buks,Levi97,AlWiMe97}. The missing $1/2$ is
recovered \cite{HRW} if one includes the Fermi sea in the QPC.

How are these results modified by the Fermi sea in the leads?
Eqs.~(\ref{scatt2}),(\ref{inter}) show that a measurement with the QPC
necessarily involves an energy transfer between the QPC and the dots. 
Our result (\ref{branching}) applies provided this transfer is not
restricted by phase space. In this case, we predict dephasing due to
the QPC even for zero temperature. This case is realized if the
applied drain source voltages are much larger than the resonance
widths as in the experiment of Ref.~\cite{Buks}. In the opposite
regime of small drain source voltages the Fermi surfaces block all 
inelastic processes. In this case, we find \cite{HRW} that dephasing
at $T=0$ is completely suppressed, in agreement with a recent general
theorem \cite{Imry}.

In summary, we have investigated the quantum Zeno effect for a system
of two quantum dots coupled to each other and to a QPC, and either
isolated from the rest of the world or connected to it by leads. In
the first case, the frequency spectrum of the QPC displays clear
signatures of the quantum Zeno effect. In the second case, the ratio
of the transmission coefficients carries similar information.

{\em Acknowledgment.} We thank Y. Imry for a valuable discussion.


\vspace{-.5cm}
\begin{references}
\vspace{-1.3cm}
\bibitem{Wheeler} J.\ A.\ Wheeler and W.\ H.\ Zurek, {\it Quantum Measurement
    Theory} (Princeton U.\ P., Princeton, 1983).

\bibitem{Zurek} W.\ Zurek, Physics Today, p.\ 36, Oct.\ 1991.

\bibitem{Misra} B.\ Misra and E.\ C.\ G.\ Sudarshan, J.\ Math.\ Phys.\ {\bf
    18}, 756 (1977). 

\bibitem{Itano} W.\ M.\ Itano, D.\ J.\ Heinzen, J.\ J.\ Bollinger, and
D.\ J.\ Wineland, Phys.\ Rev.\ A {\bf 41}, 2295 (1990). 

\bibitem{Gagen} M.\ J.\ Gagen and G.\ J.\ Milburn, Phys.\ Rev.\ A {\bf
45}, 5228 (1992).

\bibitem{Dyrting} S.\ Dyrting and M.\ J.\ Gagen, Phys.\ Rev.\ A {\bf 56}, 1453
  (1997). 

\bibitem{Gurvitz2} S.\ A.\ Gurvitz, Phys.\ Rev.\ B {\bf 56}, 15215
  (1997).

\bibitem{Buks} E.\ Buks, R.\ Schuster, M.\ Heiblum, D.\ Mahalu, and V.\
Umansky, Nature (London) {\bf 391}, 871 (1998).

\bibitem{Gurvitz1} S.\ A.\ Gurvitz, quant-ph/9607029.

\bibitem{Levi97} Y.\ Levinson, Europhys. Lett. {\bf 39}, 299 (1997).

\bibitem{AlWiMe97} I.\ L.\ Aleiner, N.\ S.\ Wingreen, and Y.\ Meir,
 Phys. Rev. Lett. {\bf 79}, 3740 (1997).

\bibitem{landau} L.\ D.\ Landau and E.\ M.\ Lifschitz, {\em Mechanics}
  (Pergammon Press, Oxford, 1969)

\bibitem{LSZ} C.\ Itzykson and J.-B.\ Zuber, {\it Quantum Field Theory} 
  (McGraw-Hill, New York, 1980).

\bibitem{HRW} G.\ Hackenbroich, B.\ Rosenow, and H.\ A.\
  Weidenm\"uller, unpublished.


\bibitem{Imry} Y.\ Imry, to be published.


\end{references}
\end{document}